\documentclass[12pt,preprint]{aastex}

\newcommand{\rosat}{{\sl ROSAT\/}}
\newcommand{\asca}{{\sl ASCA\/}}
\newcommand{\sax}{{\sl Beppo-SAX\/}}
\newcommand{\xte}{{\sl RXTE\/}}
\newcommand{\chandra}{{\sl Chandra\/}}
\newcommand{\xmm}{{\sl XMM-Newton\/}}
\newcommand{\swift}{{\sl Swift\/}}

\newcommand{\eps}{ergs\,s$^{-1}$}
\newcommand{\epcs}{ergs\,cm$^{-2}$s$^{-1}$}

\newcommand{\kms}{km\,s$^{-1}$}

\shorttitle{Novae as X-ray Transients}
\shortauthors{Mukai et al.}

\begin{document}


\title{Novae as a Class of Transient X-ray Sources}


\author{K. Mukai\altaffilmark{1,2},
        M. Orio\altaffilmark{3,4,5} and M. Della Valle\altaffilmark{5,6}}
\altaffiltext{1}{CRESST and X-ray Astrophysics Laboratory, NASA/GSFC,
        Greenbelt, MD 20771; mukai@milkyway.gsfc.nasa.gov}
\altaffiltext{2}{Department of Physics, University of Maryland,
        Baltimore County, 1000 Hilltop Circle, Baltimore, MD 21250}
\altaffiltext{3}{Department of Astronomy, University of Wisconsin,
        475 North Charter Street, Madison, WI 53706; orio@astro.wisc.edu}
\altaffiltext{4}{INAF -- Osservatorio Astronomico di Padova,
        Vicolo dell' Osservatorio 5, I-35122 Padova, Italy}
\altaffiltext{5}{Kavli Institute of Theoretical Physics, UC Santa Barbara,
        California 93106}
\altaffiltext{6}{INAF, Osservatorio Astrofisico di Arcetri, Largo Enrico
        Fermi 5, 50125 Firenze, Italy; massimo@arcetri.astro.it}



\begin{abstract}

Motivated by the recently discovered class of faint (10$^{34}$--10$^{35}$ \eps)
X-ray transients in the Galactic Center region, we investigate the 2--10 keV
properties of classical and recurrent novae.  Existing data are consistent
with the idea that all classical novae are transient X-ray sources with
durations of months to years and peak luminosities in the
10$^{34}$--10$^{35}$ \eps\ range.  This makes classical novae a viable
candidate class for the faint Galactic Center transients.  We estimate
the rate of classical novae within a 15 arcmin radius region centered on
the Galactic Center (roughly the field of view of \xmm\ observations centered
on Sgr A*) to be $\sim$0.1 per year.  Therefore, it is plausible that
some of the Galactic Center transients that have been announced to date
are unrecognized classical novae.  The continuing monitoring of the
Galactic Center region carried out by \chandra\ and \xmm\ may therefore
provide a new method to detect classical novae in this crowded and obscured
region, where optical surveys are not, and can never hope to be, effective.
Therefore, X-ray monitoring may provide the best means of testing the
completeness of the current understanding of the nova populations.

\end{abstract}




\section{Introduction}

Recently, several groups have reported their detections of relatively
faint X-ray transients in the \chandra\ and \xmm\ observations
of the Galactic Center region \citep{Pea2005,Sea2005,Mea2005}.
These authors conclude that these transients are collectively located
near the Galactic Center, based on their absorbing columns and their sky
distribution, although no direct distance measurements are available.
With this assumption, the inferred luminosities of these transients are
in the 10$^{34}$--10$^{35}$ \eps\ range.  The authors of these studies
claim that such a luminosity is too high for cataclysmic variables (CVs),
semi-detached binaries in which the accreting object is a white dwarf. 
Instead, they argue for neutron star or black hole accretors based solely
on the luminosity.  However, the Galactic Center transients are sub-luminous
compared to the known transient populations of black hole or neutron star
binaries \citep{Sea2005,Mea2005}, requiring a new population (see,
e.g., \citealt{KW2006}).

The accretion driven X-ray luminosities of CVs are indeed insufficient to
explain the Galactic Center transients.  Non-magnetic CV X-ray luminosities
are in the range 10$^{30}$--10$^{32}$ \eps, with the highest value being
3$\times 10^{32}$ \eps\ for the old nova, V603~Aql \citep{Bea2005}.
Magnetic CVs, the intermediate polars (IPs) in particular, are more
luminous in 2--10 keV X-rays, with estimated luminosities often exceeding
10$^{33}$ \eps\ \citep{SRea2006}.  However, since the highest luminosity
recorded for an IP is 1.3$\times 10^{34}$ \eps\ during the outbursts of
the unusual IP (and another old nova), GK~Per \citep{Hea2004}, IPs are
also not likely candidates for the Galactic Center X-ray transients.

However, the above discussion is incomplete because it is limited
to the accretion driven X-ray luminosities of CVs.  In reality,
CVs can generate higher X-ray luminosities through nuclear fusion,
which is a more efficient source of energy than accretion onto a white
dwarf.  Indeed, classical novae have been known to emit 2--10 keV X-rays
at luminosities exceeding 10$^{34}$ \eps.  We present below a summary of
X-ray properties of classical as well as recurrent novae.

\section{Novae as X-ray Transients}

A white dwarf accreting at below the critical rate will undergo
a thermonuclear runaway and becomes a classical nova, once a
sufficient amount of fresh fuel has been accumulated (see, e.g.,
\citealt{S1989} for a review).  A classical nova releases enough energy
($\sim 10^{45}$ ergs) to eject a shell of up to $\sim 10^{-4}$ M$_\odot$
at a typical velocity of 1000 \kms.  Classical novae are seen as
spectacular optical transients that brighten by over 10 magnitudes,
reaching peak brightness as high as M$_v = -9$ \citep{DVL1995}.  By
definition, a classical nova has only been observed to go into outburst
once, although they are thought to repeat with a recurrence period of
well over 1,000 years.  A recurrent nova is a closely related system
that has been seen to undergo multiple thermonuclear runaways; theories
of thermonuclear runaways require a high mass white dwarf accreting at
a high rate to explain the short recurrence times of these objects.

Imaging X-ray observations of classical novae weeks to months after
visual peak have revealed at least two kinds of X-ray emission \citep{O2004}.
Of these, the supersoft emission peaks in the EUV/soft X-ray range
with little or no flux above 1 keV.  We do not discuss supersoft emission
further in this paper, because supersoft emission is easily absorbed
by the interstellar medium and is unobservable from sources in
the Galactic Center region.

The other component is inferred to be from shocks within the ejected
shell, although they are spatially unresolved within the first
few years.  The X-ray spectrum of the shell component can be modeled
as optically thin thermal emission with temperatures in the 1--10 keV
range in the early stages.  The line-rich emission detected in some
novae at a later stage are also likely to be from the shell, although
they become too soft to be observable from the Galactic center region.
The review by \cite{O2004} has firmly established that some classical
novae are 2--10 keV X-ray sources.  We now attempt the first systematic
survey to see how widespread this component is, and how bright they
are on average.

\begin{figure}
\plotone{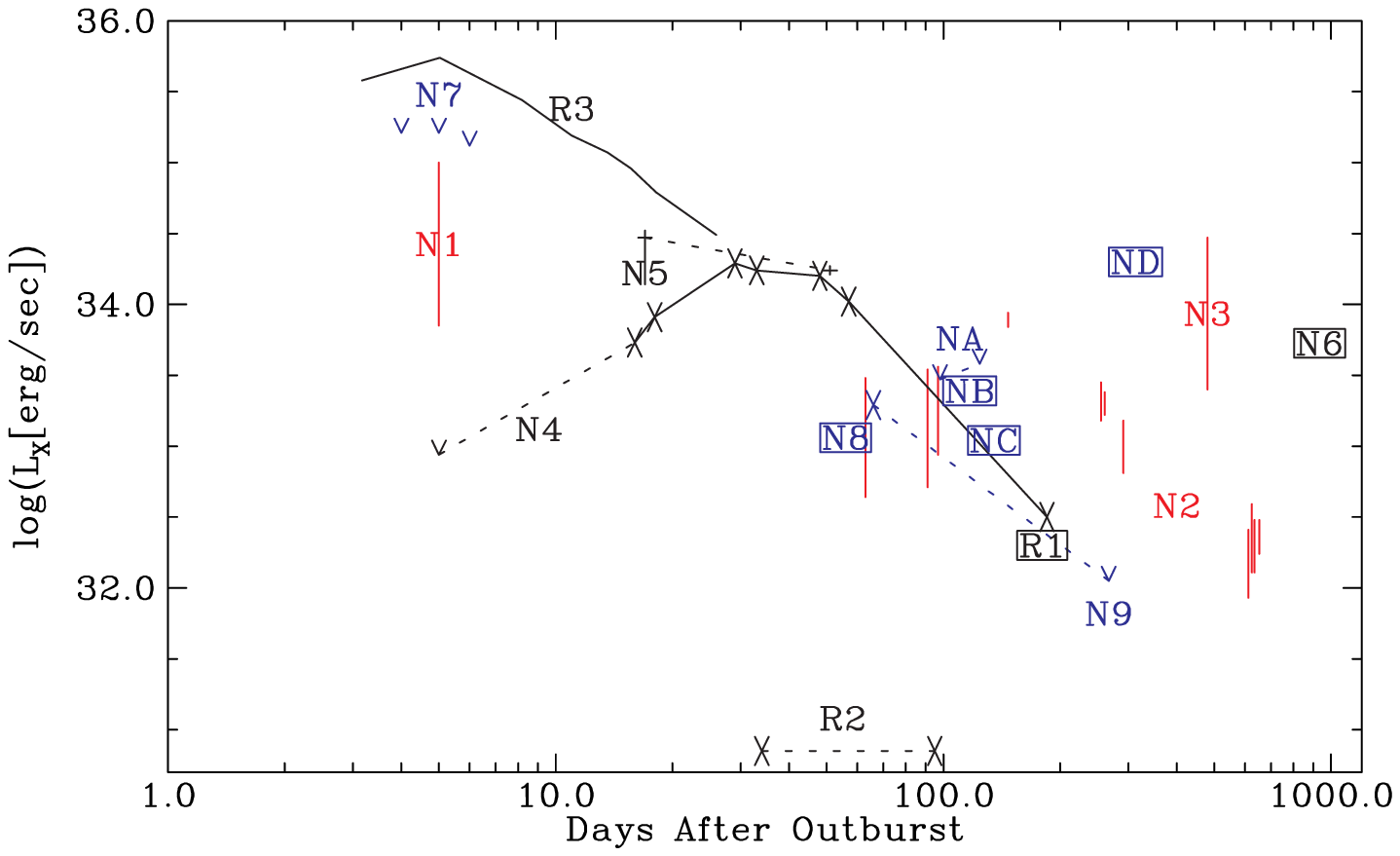}
\caption{Hard X-ray light curves of classical novae, all shown against days
since visual maximum.  Black points are generally inferred 2--10 keV
luminosities.  Blue points are the same estimated from \swift\ XRT count
rates, while red points are inferred 0.2--2.4 keV
luminosities from \rosat\ data.  Points for any given object are connected,
except that the 11 points for V1974~Cyg are left unconnected for clarity.
Upper limits are shown as upside down carets; measurements are shown using
a variety of symbols to allow those for different objects (indicated by the
object keys, see below) to be distinguished.  In 6 cases, object keys
themselves, enclosed in boxes, are used to plot measurements.  Classical
novae plotted are:
N1: V838~Her; N2: V1974~Cyg; N3: V351~Per; N4: V382~Vel; N5: Nova LMC 2000; 
N6: V4633 Sgr; N7: Nova LMC 2005; N8: V5116~Sgr; N9: V1663~Aql; NA: V1188~Sco;
NB: V477~Sct; NC: V476~Sct; ND: V382~Nor.
Recurrent novae plotted are:
R1: IM~Nor; R2: CI~Aql; R3: RS~Oph.  See text for details.}
\label{hxcurve}
\end{figure}

We present a summary in Table\,\ref{hxtab} and Figure\,\ref{hxcurve}
compiled from the literature, as detailed below.  We focus on the
first 1,000 days since eruption.  In the figure, we use
labels such as ``N1'' defined in the table.  For the three novae detected
with \rosat, we list their observed 0.2-2.4 keV luminosities and plot
them in red.  Their 2--10 keV luminosities are rather
uncertain due to the mismatch with the \rosat\ bandpass.

For other novae, we generally present the measured, absorbed 2--10 keV
luminosity.  Note that the values in Table\,\ref{hxtab} do not necessarily
coincide with those that are reported in the references, sometimes because
we have used more recent distance estimates, and because some papers 
report unabsorbed and/or bolometric luminosities.

Descriptions of the descriptions for the first
5 objects in Table\,\ref{hxtab} can be found in \cite{O2004}.
Of the other novae observed with \xmm\ or \chandra, the \xmm\ detections
of V2487~Oph (986 and 1187 days after outburst; \citealt{HS2002})
are thought to be of accretion driven X-rays, and hence we do not
include these in our summary.  More recently, V4633~Sgr has been
detected with \xmm\ (934, 1083, and 1265 days after outburst;
\citealt{HS2007}).  The authors favor a shell origin, although
cannot completely exclude accretion origin, either.  Our summary includes
only the first point (the other two, beyond our 1,000 day limit,
are at similar levels but with larger error bars).  This object is
an exception in that we have used the unabsorbed
0.2--10 keV luminosity reported by \cite{HS2007}.  The spectral
model consists of a soft and a hard component, so the 2--10 keV
luminosity should be somewhat lower.

The remainder of classical novae are taken from the \swift\ survey of
classical novae \citep{Nea2007}.  As the focus of this paper is the
supersoft component, they do not provide luminosities or the conversion
factor appropriate for the shell X-rays.  Since none of the
\swift\ observations are deep enough to enable spectroscopy of the
shell X-rays, we have used a single conversion rate of
6.24$\times 10^{-14}$ \epcs\ (2--10 keV) per 1 \swift\ XRT
cts\,ks$^{-1}$, appropriate for a kT=5 keV bremsstrahlung observed
through N$_{\rm H}$ = 1$\times 10^{22}$ cm$^{-2}$.  Among the objects
included in the \cite{Nea2007} compilation, we exclude V723~Cas,
V1494~Aql and V4743~Sgr (observed only after day 1,000); V1047~Cen
for which no distance estimate is available; and V574~Pup because
its detections are dominated by the supersoft component.

\begin{figure}
\plotone{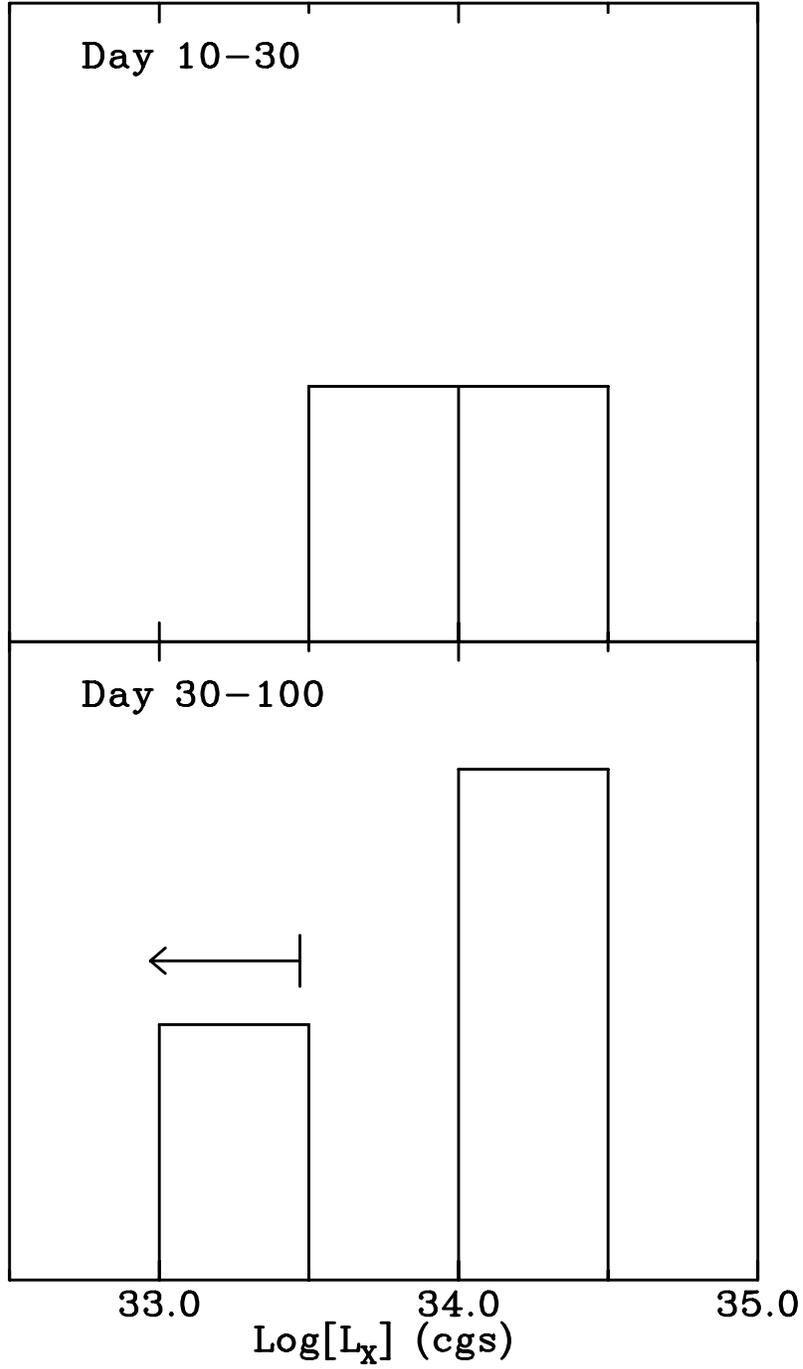}
\caption{Histograms of 2--10 keV luminosities of classical novae during
days 10-30 (top) and 30--100 (bottom).  The top panel reports 4 independent
detections of two novae, the bottom 6 detections of 4 objects and one upper
limit for a 5th system.}
\label{hxhisto}
\end{figure}

This brief summary (see also \citealt{OCO2001,O2004}) leads to the
following conclusion.  These 2--10 keV X-rays from the ejected shells
is a widespread phenomenon: of the 11 observations of 5 classical
novae in the 2--10 keV band (i.e., excluding \rosat) during day 10--100,
all but one were detections at above 10$^{33}$ \eps.  The lone exception is
the observation of V1188~Sco on day 98, with a relatively weak upper
limit.  These points are summarized in the form of two histograms
(one for day 10--30, the other for day 30--100) of 2--10 keV luminosities
of classical novae in Figure\,\ref{hxhisto}.
The existing data are consistent with the hypothesis that all
classical novae are transient 2--10 keV sources at above 10$^{33}$ \eps.
There are variations in the duration and peak luminosity
from nova to nova, but some are known to exceed 10$^{34}$ \eps.  The case
of V382~Vel suggests that there is a delay in hard X-ray turn-on of
classical novae compared to the optical peak.  

The range of
plasma temperatures in the ejecta decrease  from 20--30 keV at hard
X-ray turn-on, to $\simeq$ 1 keV in a few months (e.g., 
\citealt{Lea1992,MI2001}).  Within 1-2 years the nebula may have a rich
line spectrum, emitting mostly below 1 keV (e.g., \citealt{Nea2003,Nea2005}).
Of the novae discussed above, the time for the hard component of the X-ray
emission to cool was about 6 months for the two fast novae
\citep{BKO1998,MI2001}, but was longer (over 18 months) for slow novae with
massive ejecta \citep{Oea1996,Gea2003}, potentially exceeding the
duration of the supersoft phase.  However, the gradual decrease in
temperature means that the duration of novae as $>$2 keV X-ray sources
is effectively shorter than the total duration of novae as shell
X-ray sources.

Even less is known of the X-ray emission from recurrent novae.  IM~Nor
(R1 in Figure\,\ref{hxcurve}) was not detected 1 month after outburst
and was only a moderately strong ($\sim 2 \times 10^{32}$ [d/1 kpc]$^2$ \eps)
source 6 month past maximum \citep{Oea2005}.  The hard component of CI~Aql
(R2) was detected 34 and 95 days after outburst at about
7$\times 10^{30}$ \eps\  \citep{GdS2002} using the distance of
2.6 kpc \citep{Lea2004}.  In contrast, RS~Oph (R3) reached a luminosity
in excess of $>$10$^{35}$ \eps\ shortly after the outburst peak
\citep{Sea2006,Bea2006}. In Figure\,\ref{hxcurve}, we plot only the
observed 2--10 keV luminosity from early \swift\ observations for RS~Oph;
\xte\ measurements are similar.   The fast turn-on and high luminosity
of RS~Oph is due to the existence of an M giant wind, which provides an
additional mechanism for X-ray production not available in classical
novae or to many recurrent novae, whose mass donors are on or near the
main sequence.  The relative paucity of X-ray data on recurrent novae
reflects the fact that recurrent novae are much rarer than classical
novae.  In the rest of the paper, we will therefore concentrate on
classical novae, but the possibility of an RS~Oph-like transient
near the Galactic Center region should be kept in mind.

\section{Novae As Galactic Center Transients?}

As our summary shows, novae are a known class of X-ray transients
with peak luminosities above 10$^{34}$ \eps.  Thus, they should be
considered as a candidate class in discussing Galactic Center
transients. In fact, novae are the only known class of transients
with the right characteristics, as the known neutron star and black
hole transients have much higher peak luminosities.

Classical novae can be found both in a relatively young population
(e.g., the Galactic disk) and in the older population (e.g., the
Galactic bulge).  \cite{DVD1993} have shown (their Fig. 1) that the
distribution of the rates of decline of classical novae in the Milky
Way and in M31 perfectly overlap with each other, and both are statistically
distinguishable from LMC distribution (which exhibits a predominance of fast
rates of decline).  Since it is well known from theoretical studies
(e.g. \citealt{Sea1985,KP1985,L1992}) that the rate of decline is a tracer of
the mass of the white dwarf in the nova system, we can assume that the
main bulk of the progenitors of novae in the Milky Way and in M31
originates in the same type of stellar population.  \cite{Cea1989} and
\cite{SI2001} have demonstrated that novae in M31 are mainly produced in
the bulge (see also the tabulation of M31 novae by \citealt{Pea2007}),
therefore in view of what is reported above, the same should occur
for novae in our Galaxy.

A global Milky Way rate of $\sim$24 novae yr$^{-1}$ has been measured by
\cite{DVL1994} by scaling from extragalactic nova surveys \citep{DVea1994}.
A somewhat larger estimate of $\sim$35 novae yr$^{-1}$ has been obtained by
\cite{S1997} by extrapolating from the current rate of nova discovery
in the Galaxy (about 4--5 novae yr$^{-1}$) and by \cite{Dea2006} based on
a microlensing survey of M31.  In the following we will adopt 
as an ``educated'' guess a global rate of 30 novae yr$^{-1}$, and estimate the
rate of novae in a region of the sky within 15 arcmin of the Galactic
Center.  This is roughly the field of view of \xmm\ EPIC observations
centered on Sgr A*.

Recent estimates of the ratio nova\_rate(disk)/nova\_rate(bulge) range
from 0.25 up to 0.40 \citep{Cea1989,DVea1992,DVea1994,SI2001}.
By assuming from \cite{RvdB1989} a surface area for the Galactic disk of
850 kpc$^2$ and a typical scale height of 100 pc for disk novae
\citep{DVL1998}, the density of nova outburst in the Milky Way disk is
$\rho_{\rm disk}$=0.4--0.7 $\times 10^{-10}$ novae\,pc$^{-3}$yr$^{-1}$.
Assuming a distance from the Sun to the Galactic Center of 8 kpc,
one can find that the rate of disk novae within 15 arcmin of the Galactic
center is only $5 \times 10^{-4}$ novae yr$^{-1}$.  That is, Galactic
Center X-ray transients are highly unlikely to be disk novae.

More uncertain is the estimate of the nova density in the bulge.
Let us assume (from Figure 1 of \citealt{DVL1998}) that
most bulge novae are located within 400 pc of the Galactic
plane.  From Figure 2 of \cite{S1997} we can assume (rather
optimistically) that most bulge novae occur within the first kpc from
the Galactic center.  Under these assumptions, we find that bulge novae
are distributed within a prolate ellipsoid with a density of
$\sim 3 \times 10^{-8}$ novae pc$^{-3}$yr$^{-1}$.  The line of sight
region within 15 arcmin of the Galactic center encompasses a volume of
$\sim 35^2$ pc$^2 \times \pi \times 1000$ pc = $3.8 \times 10^6$
pc$^3$.  The expected number of bulge novae in this volume is therefore
of order $\sim$0.1 novae yr$^{-1}$.

The majority of these novae go undiscovered.  During 1978--1993, the
average rate of discovery of Milky Way novae was 3.3 yr$^{-1}$ \citep{LM1987}.
Even though the rate of discovery may have increased in recent years
(about 6 yr$^{-1}$ are reported in IAU Circulars since 2001), this still
leaves of order 25 classical novae every year that are undiscovered. 
We expect that the undiscovered novae are preferentially located in
crowded regions and/or behind high interstellar extinction.  Both problems
are extreme in the Galactic Center region.  Therefore, optical observations
are unlikely to yield a complete census of the novae in the Galactic Center
region, although wide area IR monitoring should be able to do so.

There have been observations of the Galactic Center region roughly
every 6 months with \xmm\ for roughly 2 years between 2000 Sep and
2002 Oct, out of which three transients were discovered
\citep{Pea2005,Sea2005}.  To this, we add 1 year as the representative
duration of novae as a Galactic Center X-ray transients (i.e., bright
enough and hard enough to be detectable if they were placed at the
Galactic Center; the precise value one adopts affects the following
numbers only slightly).  With this assumption, these \xmm\ observations should
have been sensitive to novae that peaked optically between 1999 Sep and
2002 Oct, or a period of 3 years.  Combined with the above estimate of
0.1 novae per year within 15 arcmin of the Galactic center, roughly the
field-of-view of \xmm\ EPIC cameras, we predict  these observations
should have detected 0.3 novae as X-ray transients.  If this is the
true expectation value, there is a 26\% chance that at least one of the
Galactic Center transients is a nova according to the Poisson distribution
(4\% chance that two or more were novae).

Most optimistically, then, one or two of the \xmm\ discovered transients
could have been unrecognized novae.  On the other hand, it may well be the
case that none of these transients are novae.  Novae are poorer candidates
for the \chandra\ transients, given the strong concentration of \chandra\ 
transients near Sgr A* \citep{Mea2005}.  However, given the uncertainties
involved both in the nova rate and the transient rate, we consider it
advisable to keep novae in mind, particularly as regular monitoring of
the Galactic Center region continues \citep{Wea2006}.

In fact, we can turn this argument around.  There is a possibility that
the present estimate of the Milky Way nova rate ($\sim$30 yr$^{-1}$) is
underestimated, because optical monitoring is ineffective in the
crowded, high extinction regions around the Galactic Center.  The
degree of central concentration of bulge novae is unknown; if there
is an additional population of novae found preferentially near the
Galactic Center, we would not know it from optical data.  The
continuing search for faint X-ray transients in the Galactic Center
region can therefore be considered an important complementary method
for discovering classical novae that are otherwise not recognized.
Since the Galactic Center region is already regularly monitored with
sensitive X-ray observatories for other purposes, it makes sense to
utilize the existing data for this purpose.

\section{Conclusions}

Classical and recurrent novae are a known class of transient X-ray sources
that reach luminosities in the 10$^{34}$ -- 10$^{35}$ \eps\ range.  The shell
X-ray phase of novae may last months to several years, although they
probably soften as they age, gradually making them less conspicuous
above 2 keV.

Novae have the right spectral and temporal characteristics to explain some
of the faint Galactic Center transients that have been detected with
\chandra\ and with \xmm\ in recent years.  If the existing literature
accurately reflects the rate of X-ray transients near the Galactic Center,
then the known population of classical (and recurrent) novae are probably
a small, but not negligible, contributor to the overall transient population.

\citet{Mea2005} have argued that dynamical processes may lead to a high
space density of X-ray binaries within 1 pc of the Galactic Center.
That is, the concentration of X-ray emitters is forcing considerations
of a new population of objects not seen elsewhere in the Galaxy.
We propose that any such studies include white dwarf binaries,
since Galactic Center specific processes could produce an additional
population of novae beyond disk and bulge novae that are currently known.
The combination of X-ray monitoring and population synthesis may represent
our best hope of obtaining a complete picture of nova populations in the
Galaxy, because optical surveys cannot possibly be effective in the
Galactic Center region.  Note that optical surveys fail to detect the
majority of Galactic novae overall --- compare the inferred Galactic nova
rate of $\sim$30 yr$^{-1}$ to the actual rate of discovery ($\sim$6 yr$^{-1}$).

\chandra\ and \xmm\ have been monitoring the Galactic Center region
more or less regularly over the last $\sim$8 years.  Even at 0.1 novae
per year within 15 arcmin of the Galactic Center, it is probable
that a nova will be detected as a 2--10 keV X-ray transient soon,
if one hasn't been already.  A concurrent infrared monitoring campaign
will be required, however, to prove beyond a reasonable doubt that
a particular X-ray transient is due to a classical nova.



\clearpage


\clearpage

\begin{deluxetable}{lrrrrrrrr}
\rotate
\tablecaption{X-ray Observations of Classical Novae in Outburst\label{xltab}}
\tablehead{\colhead{Nova} & \colhead{Dist. (kpc)} &
           \colhead{Obs.} & \colhead{Day} & Log L$_x$ (cgs)}
\startdata
 V838 Her (N1)  & 3.4 [1]   & \rosat &   5 & 33.85--35.00 [2] \\
                &           & \rosat & 370 & --\tablenotemark{a} \\
                &           & \rosat & 576 & 30.62--33.76 [3] \\
V1974 Cyg (N2)  & 1.9 [4]   & \rosat &  63 & 32.64--33.48 [5] \\
                &           & \rosat &  91 & 32.71--33.54 [5] \\
                &           & \rosat &  97 & 32.94--33.56 [5] \\
                &           & \rosat & 147 & 33.84--33.94 [5] \\
                &           & \rosat & 255 & 33.18--33.45 [5] \\
                &           & \rosat & 261 & 32.22--33.38 [5] \\
                &           & \rosat & 291 & 32.81--33.18 [5] \\
                &           & \rosat & 612 & 31.93--32.41 [5] \\
                &           & \rosat & 624 & 32.11--32.59 [5] \\
                &           & \rosat & 635 & 32.11--32.48 [5] \\
                &           & \rosat & 653 & 32.24--32.48 [5] \\
 V351 Pup (N3)  & 4.7 [6]   & \rosat & 480 & 33.40--34.47 [6] \\
 V382 Vel (N4)  & 1.7 [7]   & \xte   &   5 & $<$32.94 [8] \\
                &           & \sax   &  16 & 33.63--33.83 [9] \\
                &           & \asca  &  18 & 33.81--34.01 [8] \\
                &           & \xte   &  29 & 34.19--34.39 [8] \\
                &           & \xte   &  33 & 34.14--34.34 [8] \\
                &           & \xte   &  48 & 34.10--34.30 [8] \\
                &           & \xte   &  57 & 33.92--34.12 [8] \\
                &           & \sax   & 185 & 32.40--32.60 [9] \\
N LMC 2000 (N5) & 55        & \xmm   &  17 & 34.14--34.52 [10] \\
                &           & \xmm   &  51 & 34.21--34.27 [10] \\
                &           & \xmm   & 294 & $<$32.92 [10] \\
V4633 Sgr (N6)  & 8.9 [11]  & \xmm   & 934 & 33.60--33.85 [12] \\
N LMC 2005 (N7) & 55        & \swift &   4 & $<$35.21 [13] \\
                &           & \swift &   5 & $<$35.31 [13] \\
                &           & \swift &   6 & $<$35.12 [13] \\
V5116 Sgr (N8)  & 11.3 [13] & \swift &  56 & 32.28--33.32 [13] \\
V1663 Aql (N9)  & 5.5 [13]  & \swift &  66 & 33.08--33.43 [13] \\
                &           & \swift & 267 & $<$32.05 [13] \\
V1188 Sco (NA)  & 7.5 [13]  & \swift &  98 & $<$33.47 [13] \\
                &           & \swift & 124 & $<$33.58 [13] \\
 V477 Sct (NB)  & 11 [13]   & \swift & 117 & 33.24--33.50 [13] \\
                &           & \swift & 125 & 32.95--33.37 [13] \\ 
 V476 Sct (NC)  & 4 [13]    & \swift & 135 & $<$32.99 [13] \\
 V382 Nor (ND)  & 13.8 [13] & \swift & 313 & 34.23--34.36 [13] \\
\enddata

\tablenotetext{a}{Marginal detection but compatible with flux of the same
                  order as on day 576}

References:[1] \cite{LHR1992} [2] \cite{Lea1992} [3] \cite{SH1994}
[4] \cite{Rea1994} [5] \cite{BKO1998} [6] \cite{Oea1996} [7] \cite{DVea2002}
[8] \cite{MI2001} [9] \cite{Oea2001} [10] \cite{Gea2003} [11] \cite{Lea2001}
[12] \cite{HS2007} [13] \cite{Nea2007}

\label{hxtab}
\end{deluxetable}
 
\end{document}